\documentclass[11pt]{article}
\usepackage{latexsym}

\renewcommand{\@}[1]{\sqrt{#1}}

\def\be{\begin{eqnarray}}
\renewcommand{\le}[1]{\label{#1}\end{eqnarray}}
\def\ee{\end{eqnarray}}

\def\ffract#1#2{\raise .35 em\hbox{$\scriptstyle#1$}\kern-.25em/
\kern-.2em\lower .22 em \hbox{$\scriptstyle#2$}}

\begin{document}
\rightline{\today}
\rightline{DFTT 18/2001}
\rightline{FNT-T 16/2001}
\rightline{SISSA 56/2001/EP}
\rightline{hep-th/0107112}
\rm\large
\vskip 0.3in
\begin{center}
{\LARGE Implementing holographic projections in Ponzano--Regge gravity}
\end{center}
\vspace{1cm}
\begin{center}
{\large Giovanni Arcioni,${}^{\dagger}${}\footnote{Address after 
October 2001: Spinoza Institute, Utrecht, The Netherlands}
\footnote{
E-mail: {\tt arcioni@to.infn.it}}
Mauro Carfora,${}^\star$\footnote{E-mail: {\tt mauro.carfora@pv.infn.it}}
 Annalisa Marzuoli${}^\star$
\footnote{E-mail: {\tt annalisa.marzuoli@pv.infn.it }}
and Martin O'Loughlin${}^{\natural}$\footnote{
E-mail: {\tt loughlin@sissa.it}}}\\
\vskip 1truecm
${}^{\dagger}$ {\it Dipartimento di Fisica Teorica,
Universit$\grave{a}$ di Torino,\\
INFN, Sezione di Torino,\\
Via P. Giuria 1, I--10125 Torino, Italy}
\vskip 3.5truemm
${}^{\star}$ {\it Dipartimento di Fisica Nucleare e Teorica, 
Universit$\grave{a}$ degli Studi di Pavia,\\
INFN, Sezione di Pavia,\\
Via Bassi, 6, I--27100 Pavia, Italy}
\vskip 3.5truemm
${}^{\natural}
${\it S.I.S.S.A. Scuola Internazionale Superiore di Studi Avanzati,\\
Via Beirut 4, I--34014 Trieste, Italy}
\end{center}
\vspace{.5cm}

\begin{center}
\bf Abstract
\end{center}
\vspace{.5cm}
We consider the path-sum of Ponzano-Regge with additional 
boundary contributions in the context of the holographic 
principle of Quantum Gravity.  We calculate an holographic projection in 
which the bulk partition function goes to a semi-classical limit
while the boundary state functional remains quantum-mechanical. 
The properties of the resulting boundary theory are discussed. 
\vspace{.5cm}



\newpage

\section{Introduction}
According to the holographic principle \cite{thooft1,susskind}
Quantum Gravity with some prescribed asymptotic behavior can
be described by a theory defined on the boundary at infinity.\\
Two recent examples which display holographic behavior are Matrix Theory 
\cite{bfss} and the AdS/CFT correspondence \cite{malda}. In particular 
regimes and by
considering suitable decoupling limits it has been possible to carry out
many checks of such a behavior and they all supported the holographic 
nature of theories 
containing gravity (for an exhaustive review see \cite{review}). 
This represents a new perspective which seems to 
indicate that spacetime physics emerges from an underlying theory living 
in lower dimensions.\\
The holographic principle has been originally proposed by 't Hooft 
\cite{thooft1} in discussing spacetimes containing black holes. In 
particular, the fact that the black hole
entropy is proportional to an area supports the holographic principle and the
black hole horizon is then interpreted as a screen encoding bulk information. 
This lead 't Hooft to fomulate an S-matrix Ansatz program so as to
properly incorporate black holes when dealing with quantized gravity. 
Since black holes carry the strongest possible gravitational field in any 
given volume 
they form a very natural
 upper bound of the energy spectrum.\\
A very interesting feature of the 't Hooft picture is that the theory
living on the horizon--screen  of the black hole is supposed 
to be some sort of 
discretized model \cite{thooft2} which comes about when
including the transverse gravitational force in the construction of the
S-matrix. According to preliminary attempts to implement such a picture, 
the operators
on the discretized horizon satisfy an angular momentum algebra and 
the horizon is divided into various domains representing the various 
representations
of the algebra. In the 2+1 dimensional case \cite{thooft3}
the horizon is described as being built up from string bits with unit lengths 
and a covariant algebra of observables associated to ingoing and outgoing
radiation can be indeed constructed.\\
In such a framework it seems particularly interesting to examine 
discretized models in 
the presence of a boundary as a possible test of the holographic description of
spacetime. One of the simplest and most natural scenarios is to consider the
Ponzano--Regge model in three dimensions in the presence of a boundary. For 
a general introduction to discretized gravity and to the Ponzano-Regge model 
in particular see the review \cite{rwilliams}. 
We will work in the Euclidean sector and as a preliminary investigation 
we examine
 the possibility of implementing a holographic description in the presence 
of a boundary. More precisely, we will try to implement some sort of 
holographic 
projection of bulk data precisely in the spirit of 't Hooft ideas.\\
In this connection it is worthwhile stressing that the continuum counterpart 
of the Ponzano--Regge model is three--dimensional 
Chern--Simons theory \cite{KNS}
and it is well known that for gauge group 
$SL(2,R)\times SL(2,R)$ this latter induces on the boundary a WZW model 
and  eventually Liouville field theory. This sort of "intrinsic" holographic 
behaviour has also led to the conjecture that a suitable higher dimensional 
Chern--Simons theory with some prescribed symmetries
 can be a possible candidate for a holographic description of M-theory 
\cite{horava}. Chern--Simons theory has indeed a very big symmetry group 
resembling gravity diffeomorphism invariance and this should lead to a 
holographic behavior when putting a suitably conditioned boundary in the 
theory. In view of these remarks it seems therefore interesting to analyze 
the presence of a boundary in a discretized version of Chern--Simons 
theory starting precisely from the three dimensional case.\\
The paper is organized as follows: in Section 2 we briefly recall some 
aspects of the 
Ponzano--Regge model in the presence of boundary. In Section 3
we separate the path-sum into contributions from the 
bulk, intermediate contributions that ``communicate from bulk to boundary'' 
and purely boundary terms. We then take the semi-classical limit of 
the bulk part of the path-sum and discuss the resulting action. In Section
4 we make some comments on the resulting split path-sum and the 
significance of the various terms. 
\section{Ponzano--Regge gravity in the presence of boundaries}
The original Ponzano--Regge model \cite{pr}, associated with the partition 
function of
$3$--dimensional Euclidean gravity for a closed simplicial manifold $M^3$,  
can be
extended to deal with manifolds with a non--empty boundary 
$(M^3, \partial M^3)$ 
(where $\partial M^3$ could be the disjoint union of a finite number of 
components).
The basic idea is to consider the triangulation induced on the boundary 
by the bulk simplicial
decomposition of the three-manifold since the tetrahedra that have one 
or more of their 
triangular faces in common with the boundary induce a boundary triangulation. 
The weight of the path sum associated with this boundary triangulation can 
be determined in a variety of ways. Via the canonical formalism whereby 
one considers a two-manifold that is evolving in Euclidean time \cite{ooguri}. 
Otherwise we can use a general topological construction 
\cite{carbone1,carbone2}, 
in which one enforces invariance of the path sum including boundary 
under the discretized version of diffeomorphisms \cite{pachner2}. Finally, 
via the discretization of the corresponding BF (first-order formalism for 
three-dimensional gravity) theory, imposing suitable 
boundary conditions on the path integral \cite{martin}. In each of 
these cases, the path sum includes sum over the boundary spins, and thus 
the boundary geometry has identical weights. In the case of \cite{ooguri} and 
\cite{martin}, this boundary sum remains also a function of the connection
since boundary conditions were chosen such that the connection was fixed 
on the boundary.\\ 
We give in the rest of this section a brief review of these results for the 
path sum with fluctuating boundary.
For simplicity we will take the connection to be trivial, although
in general a global gauge choice of this type cannot be made due to 
possible non--trivial holonomy along one--cycles in the boundary two--fold.\\
Let $(M^3, \partial M^3)$ be a Piecewise Linear ($PL$) pair of a fixed
topological type, and consider a particular triangulation 
$(T^3(J), \partial T^3(j;m))$ with edges,
faces and tetrahedra labelled by suitable elements of the 
Racah--Wigner algebra of
$SU(2)$ (or of its $q$-deformed counterpart, $q$ a root of 
unity \cite{turaev}).
More precisely, we weight the contribution of any such 
triangulation to the partition
function (or state sum) $Z[(M^3, \partial M^3)]$ 
by introducing the following functional \cite{carbone1}:

\begin{eqnarray}
\lefteqn{Z[(T^3(J),\partial T^3(j;m)) \rightarrow (M^3, \partial M^3); L]\equiv
Z[(T^3,\partial T^3);L]
=}\hspace{.5in}\nonumber \\
& & \Lambda(L)^{-N_0}\,\prod_{1}^{N_1} (-1)^{2J} (2J+1)\,\prod_{1}^{N_3} 
(-1)^{\sum J}\:\ \left\{6j\right\}(J)  \nonumber \\
& & \Lambda(L)^{-n_0}\,\prod_{1}^{n_1} (-1)^{2j} (2j+1)\,\prod_{1}^{N^F_3} 
(-1)^{\sum (J+j)}\:\, \left\{6j\right\}(J,j)  \nonumber \\
& & \prod_{1}^{n_2} (-1)^{(\sum m)/2}\:\, \left[3jm\right]. 
\label{tristate}
\end{eqnarray}

\noindent Here $N_0,N_1,N_3$ denote the number of vertices, edges and 
tetrahedra which
lie entirely in the interior of $T^3$ ($Int T^3$), while $n_0,n_1,n_2$ 
are the number of
vertices, edges and triangular faces in $\partial T^3$; $N^F_3$ is the 
number of tetrahedra
which have faces in $\partial T^3$.\\
The set of spin variables $\{J\}$ labels edges in $Int T^3$, while the 
set $\{j\}$ is associated
with edges in $\partial T^3$; each spin variable runs over 
$\{0,1/2,1,\ldots\}$.
The cut--off $L$ plays here the same role as in Ponzano--Regge 
asymptotic formula \cite{pr},
{\em i.e.} $\Lambda(L)=4L^3/3c$ ($c$ an arbitrary constant), and 
all spin variables in (\ref{tristate})
are bounded above by $L$.\\
Each of the symbol in (\ref{tristate}) has a precise group theoretical 
and geometrical meaning,
namely:

\begin{itemize}

\item $(2J+1)$ [respectively $(2j+1)$], the dimension of the $J$--th 
[respectively 
$j$--th] irreducible representation of
$SU(2)$, gives the contribution to the measure of each edge belonging to 
$IntT^3$ [respectively $\partial T^3$].

\item The Racah--Wigner $6j$ symbol \par

\begin{equation}
\left\{6j\right\}(J)\,\equiv\,
\left\{ \begin{array}{ccc}
J_1 & J_2 & J_3 \\
J_4 & J_5 & J_6
\end{array}\right\}
\label{Jtetra}
\end{equation}

\noindent represents, apart from the phase factor $(-1)^{\sum J}$, 
$\sum J=J_1+J_2+\ldots+J_6$,
a tetrahedron in $Int T^3$.\\
The notation $\left\{6j\right\}(J,j)$ stands for a tetrahedron which 
has some of its faces in
$\partial T^3$, and thus the entries $\{J,j\}$ of such symbol depend on 
its place in the
triangulation. 

\item The Wigner $3jm$ symbol \par

\begin{equation}
\left[3jm\right]\,\equiv\,
\left[ \begin{array}{ccc}
j_1 & j_2 & j_3 \\
m_1 & m_2 & -m_3
\end{array}\right],
\label{jtri}
\end{equation}

\noindent apart from the phase factor $(-1)^{(\sum m)/2}$, 
$\sum m =m_1+m_2+m_3$, is
associated with a triangular face in $\partial T^3$, and each $m$-variable 
(representing the
projection of the corresponding $j$ with respect to a fixed 
quantization axis) satisfies
the usual requirement $-j\leq m \leq j$ in integer steps.

\end{itemize}

\noindent
The partition function to be associated with the manifold 
$(M^3, \partial M^3)$ is found
by summing 
the contributions of the type (\ref{tristate}) over all admissible 
angular momentum configurations for the simplicial decomposition
$(T^3(J), \partial T^3(j;m))\rightarrow (M^3, \partial M^3)$, under 
the condition that all spin variables $\{J,j\}$
-- together with the $m$'s -- are bounded above by a fixed $L$. In the 
spirit of the
Ponzano--Regge approach, in order to remove the cut--off we formally take
the limit $L \rightarrow \infty$ of the state sum, and thus the 
expression of the ``regularized'' 
partition function can be formally written as:

\begin{equation}
Z[(M^3, \partial M^3)]\,=
\,\lim_{L\rightarrow \infty}\:
\sum_
{\left \{\begin{array}{c}
(T^3, \partial T^3)\\ 
J,j,m \leq L
\end{array}\right\}}
Z[(T^3,\partial T^3) \rightarrow (M^3, \partial M^3); L].
\label{origsum}
\end{equation} 

\noindent This state sum reduces to the Ponzano--Regge functional in the case
$\partial M^3=\emptyset$, while we would recover the expression found in
\cite{ooguri} if we keep fixed the triangulation $\overline{\partial T^3}$ on
the boundary manifold, {\em i.e.} for 
$(M^3,\partial M^3=\overline{\partial T^3})$.\\
In the asymptotical limit (all spin variables large) (\ref{origsum}) 
represents the semiclassical
partition function of Euclidean gravity in the presence of a boundary 
with the action 
discretized according to the Regge prescription \cite{regge}, 
({\em cfr.} \cite{haso} for the extension of Regge
Calculus to manifolds with boundary). More precisely, for a 
triangulation $(T^3, \partial T^3)$
the action contains the Einstein--Regge action of the bulk, 
namely ${\cal I} (Int T^3)$ $\propto\,
\sum_{J}\; (J+1/2)\, \theta$ (where $\theta$'s are the deficit angles), 
and a suitable contribution
due to the boundary $\partial T^3$.\\
To discuss the role of the boundary terms in the partition
function (\ref{origsum}) we refer now to \cite{carbone3} 
(see also \cite{martin,carbone2,ooguri}), 
where the state sum model induced on 
the closed boundary manifold $\partial M^3 \equiv M^2$ was derived. 
It turns out that each triangular
face in a triangulation $T^2(j;m,m')\rightarrow M^2$ has to be 
associated with the following product of two
$3jm$ symbols

\begin{equation}
(-1)^{\sum_{s=1}^3 (m_s+m'_s)/2}\:
\left[ \begin{array}{ccc}
j_1 & j_2 & j_3 \\
m_1 & m_2 & -m_3
\end{array}\right]
\left[ \begin{array}{ccc}
j_1 & j_2 & j_3 \\
m'_1 & m'_2 & -m'_3
\end{array}\right].
\label{jmmtri}
\end{equation}

\noindent Here $\{m_s\}$ and $\{m'_s\}$ are two different sets of momentum 
projections associated with the
same angular momentum variables $\{j_s\}$, and 
$-j_s \leq m_s, m'_s \leq j_s$ $\forall 
s=1,2,3$.\\
The expression of the functional associated with a particular 
triangulation $T^2(j;m,m')$ is

\begin{eqnarray}
\lefteqn{Z[T^2(j;m,m')\rightarrow M^2; L]\,\equiv\, Z[T^2; L]\,
=}\hspace{1in} \nonumber\\
& & \Lambda(L)^{-n_0}\,\prod_{A=1}^{n_1} (2j_A+1) 
(-1)^{2j_A}(-1)^{-m_A-m'_A}\nonumber\\
& & \prod_{B=1}^{n_2} 
\left[ \begin{array}{ccc}
j_1 & j_2 & j_3 \\
m_1 & m_2 & -m_3
\end{array}\right]_B
\left[ \begin{array}{ccc}
j_1 & j_2 & j_3 \\
m'_1 & m'_2 & -m'_3
\end{array}\right]_B,
\label{bistate}
\end{eqnarray}
\noindent where $n_0, n_1, n_2$ are the numbers of vertices, edges and 
triangles 
in $T^2(j;m,m')$, respectively. Summing over all admissible triangulations 
we get the partition function of the
closed $2$-dimensional theory which reads

\begin{equation}
Z[M^2]\,=\:\lim_{L\rightarrow \infty}\:\:\sum_{\{T^2;j,m,m' \leq 
L\}}\:\:Z[T^2; L],
\label{bisum}
\end{equation}

\noindent where the regularization is carried out according to the previous 
prescription. This 
expression can be evaluated in a staightforward way, providing the expression 

\begin{equation}
Z[M^2]\,=\,\Lambda(L)\;\Lambda(L)^{-{\bf \chi}(M^2)},
\label{eulero}
\end{equation}

\noindent where ${\bf \chi}(M^2)$ is the Euler character of the manifold 
$M^2$.\\
Thus the partition function of the $2$-dimensional closed model induced 
on the boundary by 
(\ref{origsum}) contains the only topological invariant 
which is significant in the present context (recall also that the Regge action
in dimension $2$ is just ${\bf \chi}(M^2)$).\\
Note that the regularization 
prescription used above is ill defined in general as there is no natural 
way to remove the cut-off. 
The naturally regularized counterpart of the closed Ponzano--Regge model 
is the 
quantum invariant introduced in \cite{turaev} which turns out to be 
related to a double
Chern--Simons--Witten theory at level $k$ where $exp(\pi i/k)=q$ is  
the deformation parameter.
Moreover, the limit for $q \rightarrow 1$ of the Turaev--Viro 
functional corresponds to a semiclassical 
partition function containing an Einstein--Regge term plus a 
volume term with a positive
cosmological constant related to $k$ \cite{mizo} ({\em cfr.} also 
\cite{carlip,jan} on these issues).
In \cite{carbone1}, \cite{carbone2} the $q$-deformed counterparts of 
the state sums (\ref{origsum}) and
(\ref{eulero}) have been defined, and in particular (\ref{eulero}) 
corresponds to the quantum invariant

\begin{equation}
Z[M^2]_q\,=\,w_q^2\;w_q^{-2{\bf \chi}(M^2)},
\label{qeuler}
\end{equation}

\noindent where $w_q^2=(-2k)/(q-q^{-1})^2$.\\

\noindent The state sums associated with simplicial dissections of 
$3$-manifolds based
on the recoupling theory of quantum angular momenta have a rich 
topological structure which 
is non trivially implemented by exploiting a suitable set of 
operations on simplices 
(topological {\em moves}).
In general, given two compact $d$-dimensional simplicial 
$PL$--manifolds (with or
without boundaries) Pachner proved that they are $PL$--homeomorphic 
if and only if
their underlying triangulations are related to each other either by a 
finite set of {\em bistellar moves}
(in the closed case, \cite{pachner1}) or by a finite set of 
{\em elementary shellings}
(in the case with boundary, \cite{pachner2}). Since in $d\leq3$ the 
$PL$--type and the 
topological equivalence class of any compact manifold are in $1-1$ 
correspondence, in the present context 
we can simply speak of "topological equivalent" or "homeomorphic" manifolds.\\
The next step consists in recognizing that all the moves we are 
referring to can be translated 
into algebraic identities which are encoded in the structure of 
the partition functions 
written above.\\
Recall that in the closed $3$-dimensional 
case (see \cite{pr}, \cite{turaev})
the bistellar moves can be expressed algebraically in terms of the 
Biedenharn--Elliott 
identity (representing the moves 
[$2$ tetrahedra joined along a common face] $\leftrightarrow$ 
[$3$ tetrahedra joined along a common edge])
and of both the B-E identity and the 
orthogonality conditions for $6j$ symbols, which represent the barycentic move 
together with its inverse, namely [$1$ tetrahedron] $\leftrightarrow$ [$4$ 
tetrahedra]. Thus the Ponzano--Regge functional, namely 
(\ref{origsum}) for $\partial M^3=
\emptyset$, is formally a topological invariant (and the Turaev--Viro state sum
is in fact a well--defined topological invariant).\\
In \cite{carbone3} the identities representing the bistellar 
moves in $d=2$ have been established
(and they will be given explicitly in the next section). 
They are associated with the flip move
[$2$ triangles] $\rightarrow$ [$2$ triangles] 
(a pair of triangles joined to form a 
quadrilateral are transformed into two triangles joined along the 
other diagonal)
and with the barycentric move and its inverse, namely [$1$ triangle] 
$\leftrightarrow$
[$3$ triangles]. Then the fact that (\ref{bisum}) is actually a 
topological invariant
can be recognized on the basis of its manifest invariance under any 
finite set of
such moves.\\
Turning now to the case of $(M^3, \partial M^3)$, the topological 
transformations 
that have to be taken into account are the elementary shellings 
and their inverse moves
introduced in \cite{pachner2}. These operations 
involve the cancellation of one tetrahedron at a time 
from a given triangulation 
$(T^3(J), \partial T^3(j;m)) \rightarrow$ $(M^3, \partial M^3)$. 
In order to be 
deleted, the tetrahedron must have some of its $2$-dimensional faces
lying on the boundary $\partial T^3$.  Moreover, for each elementary shelling 
there exists an inverse move which corresponds to the attachment of a new 
tetrahedron to a suitable component in $\partial T^3$. In \cite{carbone1}
the identities corresponding to such moves have been found, and the 
expression of the
state sum given in (\ref{origsum}) and (\ref{tristate}) is actually manifestly 
invariant under any finite set of such moves. The conclusion is that, owing to
Pachner's results, (\ref{origsum}) is a topological ($PL$) invariant of
the pair $(M^3, \partial M^3)$. Notice also that, since 
$Z[(M^3, \partial M^3)]$
reduces to the Ponzano--Regge partition function in the case $\partial M^3=
\emptyset$, it is automatically invariant also under bistellar moves 
in the bulk $Int M^3$.\\

\section{Projecting on the boundary and the decoupling limit}

In the spirit of the holographic principle we would like
to decouple -- or more properly disentangle -- the theory living
on the boundary from the bulk gravity theory. We therefore look for some way
of projecting the theory described in the previous section onto the boundary.
Notice that the $2$-dimensional state sum given in (\ref{bisum}) 
is not generated in this way since it simply represents 
a theory obtained from the $3$-dimensional functional (\ref{tristate}) 
by restricting it to the boundary and by summing freely over all 
triangulations of 
the surface $\partial M^3=$ $M^2$, up to 
regularization.\\
To set up a true decoupling procedure one has to examine more carefully
the structure of the functional (\ref{tristate}) describing the state 
of a single
triangulation $(T^3(J), \partial T^3(j;m))$. In particular,
one expects to recognize bulk pieces, boundary pieces and interaction
terms bulk--boundary. Bulk pieces are going to describe the 
fluctuations of the
spins $J$'s in the interior of the manifold, boundary pieces the
fluctuations of the spins $j$'s  (and of their corresponding $m$'s) 
on the boundary, and the interaction terms
the "coupling" between such bulk-boundary fluctuations.\\ 
The interaction bulk--boundary is clearly given by those tetrahedra which
have some of their components on the boundary triangulation, and in 
particular by terms
of the type $\{6j\}(J,j)$ in (\ref{tristate}). Such symbols may have 
in principle
a varying number of spins of type $J$ or $j$. But, as discussed at the 
end of the previous
section, we can take advantage of 
the invariance of $Z[(M^3, \partial M^3)]$ in (\ref{origsum})
under elementary shellings. In particular we can always transform a 
given triangulation 
$(T^3(J), \partial T^3(j;m))$ $\rightarrow$ $(M^3, \partial M^3)$ in 
such a way that 
each tetrahedron which shares with the boundary a number of faces
has actually only one face in $\partial T^3$.
Call the resulting dissection {\em standard triangulation} to be
denoted from now on by
$(T^3,\partial T^3)_{st}$.\\
Note however that in a standard triangulation there are two distinct 
types of tetrahedra
having edges in $(\partial T^3)_{st}$:

\begin{itemize}
\item The tetrahedra $\{\sigma_F\}$ which are involved in the definition
of the standard triangulation itself, namely the ones which share with 
the boundary one of
their faces and the corresponding three edges. Their number $N_3^F$ is 
equal to the number of triangular faces in 
$(\partial T^3)_{st}$, which have been denoted by $n_2$ in 
(\ref{tristate}), namely we have
\begin{equation}
N_3^F\,=\,n_2.
\label{n3}
\end{equation} 

\item The tetrahedra $\{\sigma_E\}$ which have exactly one edge in 
$(\partial T^3)_{st}$
and no corresponding face (there could be a varying number of such 
tetrahedra for any 
edge in $(\partial T^3)_{st}$). Denote their total number by $N_3^E$.
\end{itemize}

\noindent We shall call the $\{\sigma_F,\,\sigma_E\}$ 
{\em coupling tetrahedra}.\\
The $\{6j\}(J,j)$ symbols to be associated with these types of 
coupling tetrahedra in the state 
functional of a standard triangulation can be always cast -- 
by making use of the symmetry
properties of the $6j$ symbols -- in two particular forms

\begin{equation}
\sigma_F \,\longleftrightarrow\,
(-1)^{\sum J + \sum j}\:\:
\left\{ \begin{array}{ccc}
j_1 & j_2 & j_3 \\
J_1 & J_2 & J_3
\end{array}\right\}\,\doteq\,(-1)^{\sum J+\sum j}\:\{6j\}(\sigma_F),
\label{Jjtetra}
\end{equation}

\noindent where the phase factor is explicitly given by 
$\sum_{i=1}^{3} J_i + \sum_{i=1}^{3} j_i$, and

\begin{equation}
\sigma_E \,\longleftrightarrow\,
(-1)^{\sum J + j}\:\:
\left\{ \begin{array}{ccc}
J_1 & j & J_2 \\
J_3 & J_4 & J_5
\end{array}\right\}\,\doteq\,(-1)^{\sum J +j}\:\{6j\}(\sigma_E),
\label{Jjtetra2}
\end{equation}

\noindent with the phase factor given by $\sum_{i=1}^{5} J_i$.\\
Thus, if we denote by

\begin{eqnarray}
N_1^F&=&\mbox{number of edges of type} \:J\: \mbox {in} \:\{\sigma_F\}\\
N_1^E&=&\mbox{number of edges of type} \:J\: \mbox {in} \:\{\sigma_E\}
\label{N1cou}
\end{eqnarray}

\noindent we can rewrite the state functional (\ref{tristate})
for a standard 
triangulation $(T^3,\partial T^3)_{st}$  by setting
 
\begin{eqnarray}
\lefteqn{Z[(T^3(J),\partial T^3(j;m))_{st}
;L]\,\equiv Z[(T^3,\partial T^3)_{st};L]
=}\nonumber \\
& & \Lambda(L)^{-N_0}\,\prod_{1}^{N_1 - N_1^{F*} - N_1^{E*}} 
(-1)^{2J} (2J+1)\,\prod_{1}^{N_3-N_3^E} 
(-1)^{\sum J}\:\ \left\{6j\right\}(J)\nonumber\\
& & \prod_{1}^{N_1^{F*}} (-1)^{2J^F} (2J^F+1)\: \prod_{1}^{N_1^{E*}} 
(-1)^{2J^E} (2J^E+1)\nonumber\\
& & \Lambda(L)^{-n_0}\:\:\prod_{1}^{n_1} (-1)^{2j}\: (2j+1)  \nonumber \\
& &  \prod_{1}^{n_2} \;
(-1)^{\sum J+ \sum j}\:\, \left\{6j\right\}(\sigma_F)\:\,
(-1)^{(\sum m)/2}\:\, \left[3jm\right]\nonumber\\
& & \prod_{1}^{N_3^E} \;
(-1)^{\sum J+  j}\:\, \left\{6j\right\}(\sigma_E).
\label{standstate}
\end{eqnarray}

\noindent where the edges of the coupling tetrahedra $\{\sigma_F,\sigma_E\}$ 
have been labelled by $F$ and $E$, respectively. 
We have introduced $N_1^{F*} < N_1^F$ and $N_1^{E*} < N_1^E$ as there 
are only $N_1^{F*} + N_1^{E*}$ independent edges of type J in 
$\{\sigma_F,\sigma_E\}$ due to the obvious identification of edges in common
to more than one of these coupling tetrahedra. \\
\noindent Ponzano--Regge gravity is a discretized counterpart of a 
second order theory in 
which the variables associated with the edges play the role of the 
metric tensor \cite{regge}.
Thus, in the spirit of an holographic
scenario, one has to take a suitable limit on some "metric variables" living
close to the boundary
in order to be able to decouple the boundary theory from the bulk.
The collection of the coupling tetrahedra $\{\sigma_F,\sigma_E\}$ is 
obviously a natural candidate to play this role. 
However, we easily recognize that there is a further class of tetrahedra 
living {\em close} to the boundary,
namely those tetrahedra with edges in $Int T^3$, but which have one of 
their
vertices in $(\partial T^3)_{st}$ (the weights of these vertices are 
included in the factor
$\Lambda(L)^{-n_0}$ of (\ref{standstate})). Denote this new 
collection by $\{\sigma_V\}$ and 
call them {\em coupling tetrahedra} too.
Their number runs, say, from $1$ to $N_3^V$, and each of them 
corresponds to a $6j$
symbol of the form

\begin{equation}
\sigma_V \,\longleftrightarrow\,
(-1)^{\sum J}\:\:
\left\{ \begin{array}{ccc}
J_1 & J_2 & J_3 \\
J_4 & J_5 & J_6
\end{array}\right\}\,\doteq\,(-1)^{\sum J}\:\{6j\}(\sigma_V),
\label{Jtetra3}
\end{equation}

\noindent where no confusion should arise with the generic symbol 
introduced in (\ref{Jtetra}).
Then, if we denote by 
\begin{equation}
N_1^V\,=\,\mbox{number of edges in}\:\{\sigma_V\}
\label{N1cou2}
\end{equation}

\noindent the first two groups of terms in the right--hand side of 
(\ref{standstate}) can be rewritten as

\begin{eqnarray}
\lefteqn{
\Lambda(L)^{-N_0}\,\prod_{1}^{N_1-N_1^{F*}-N_1^{E*}-N_1^{V*}} 
(-1)^{2J} (2J+1)\,
\prod_{1}^{N_3-N_3^E-N_3^V} 
(-1)^{\sum J}\:\ \left\{6j\right\}(J)}\nonumber\\
& & \prod_{1}^{N_1^{F*}}(-1)^{2J^F} (2J^F+1)
\prod_{1}^{N_1^{E*}}(-1)^{2J^E} (2J^E+1)
\prod_{1}^{N_1^{V*}}(-1)^{2J^V} (2J^V+1)\doteq\nonumber\\
& & \Lambda(L)^{-N_0}\,\prod_{1}^{N_1-N_1^*} (-1)^{2J} (2J+1)\,
\prod_{1}^{N_3-N_3^E-N_3^V} 
(-1)^{\sum J}\:\ \left\{6j\right\}(J)\nonumber\\
& & \prod_{1}^{N_1^*}(-1)^{2J} (2J+1),
\label{volst}
\end{eqnarray}

\noindent where $N_1^{V*}$ again takes into account the additional 
necessary identifications of edges between the tetrahedra $\sigma_V$
and the tetrahedrae $\{\sigma_F,\sigma_E\}$. 
On the right--hand side of (\ref{volst}) we have then defined

\begin{equation}
N^*_1\:\doteq \: \mbox{number of edges in}\: (Int T^3)_{st}\; \setminus 
(\{\sigma_F\}\cup\{\sigma_E\}\cup\{\sigma_V\}).
\label{N1redef}
\end{equation}

\noindent The role of the weights of such edges, labelled again by 
$J$ to simplify the notation, will be discussed below.\\
In order to set up a decoupling procedure in (\ref{standstate}) 
(taking into account 
(\ref{volst})), we have to
keep fixed all labels $\{j,m\}$ together with the $N_1-N_1^*$ 
variables of type $J$, while rescaling
the $N_1^*$ $J$-variables belonging to the coupling tetrahedra 
by {\em the same factor $R$}. More precisely,
we have different asymptotic expressions for each type of coupling tetrahedra
(see \cite{pr}, \cite{russi} and the original references therein), namely

\begin{itemize}
\item For a tetrahedron $\sigma_F$ we get

\begin{equation}
\{6j\}(\sigma_F,R)\equiv
\left\{ \begin{array}{ccc}
j_1 & j_2 & j_3 \\
J_1+R & J_2+R & J_3+R
\end{array}\right\}\:\:\:
\begin{array}{c}
\longrightarrow \\
R \gg 1
\end{array}\:
\frac{(-1)^{\Phi}}{\sqrt {2R}}\:
\left[ \begin{array}{ccc}
j_1 & j_2 & j_3 \\
\mu_1 & \mu_2 & \mu_3
\end{array}\right],
\label{astetra}
\end{equation}

\noindent where $\Phi=j_1+j_2+j_3+2(J_1+J_2+J_3)$. Such an expression 
tells us that by rescaling
the three $J$-variables by a large $R$, this $6j$ symbol
goes into a weighted Wigner $3jm$ symbol with the same $j$'s, 
and with momentum projections depending
on differences of $J$'s according to

\begin{equation}
\left\{
\begin{array}{ccc}
\mu_1 & = & J_2-J_3, \\
\mu_2 & = & J_3-J_1, \\
\mu_3 & = & J_1-J_2. 
\end{array}\right.
\label{muJ}
\end{equation}

\item For a tetrahedron $\sigma_E$ we get

\begin{equation}
\{6j\}(\sigma_E,R)\equiv
\left\{ \begin{array}{ccc}
J_1+R & j & J_2+R \\
J_3+R & J_4+R & J_5+R
\end{array}\right\}\:\:\:
\begin{array}{c}
\longrightarrow \\
R \gg 1
\end{array}\:
\frac{(-1)^{\Psi}}{2R}\,\:
d^j_{\nu_2\nu_3}(\theta),
\label{astetra1}
\end{equation}

\noindent where $\Psi=3J_1+j+2(J_2+J_3+J_4+J_5)+\nu_1$ and $\theta$ 
is the angle between the
the edge labelled by $j$ and the quantization axis. Thus we get that 
the asymptotic behavior
of such a symbol goes like a product of a Kronecker $\delta$ and a 
Wigner $d$--function 
with entries given by

\begin{equation}
\left\{
\begin{array}{ccc}
\nu_1 & = & J_3-J_4, \\
\nu_2 & = & J_5-J_3, \\
\nu_3 & = & J_2-J_1. 
\end{array}\right.
\label{nuJ}
\end{equation}

\item Finally, the asymptotic expression for a tetrahedron $\sigma_V$ 
is nothing but the original
Ponzano--Regge formula \cite{pr}, namely

\begin{eqnarray}
\{6j\}(\sigma_V,R)\equiv
\left\{ \begin{array}{ccc}
J_1+R & J_2+R & J_2+R \\
J_4+R & J_5+R & J_6+R
\end{array}\right\}\:\:\:
\begin{array}{c}
\longrightarrow \\
R \gg 1
\end{array}\nonumber\\
\left( 12\pi {\cal V}(\sigma_V)\right)^{-1/2}\;
\exp \left\{\; i\left( \; \sum_{\alpha=1}^{6}l_{\alpha}\theta_{\alpha} + 
\pi/4 \right) \right\}. 
\label{asPR}
\end{eqnarray}

${\cal V}(\sigma_V)$ is the Euclidean volume of the tetrahedron  
spanned by the six edges
$\{l_{\alpha}\}$, $l_{\alpha}=J_{\alpha}+1/2$ and $\theta_{\alpha}$ 
is the 
angle between the outward normals to the faces
which share $l_{\alpha}$ (these angles can be obviously expressed in 
terms of the $J$'s).
Note that the sum under the exponential is the Regge action for the 
tetrahedron $\sigma_V$.\\

\end{itemize}

\noindent Since we can freely choose the decoupling parameter 
$R\gg J (\leq L)$ at this level,
we carry out the rescaling both on the three types of $6j$ symbols 
according to 
(\ref{astetra}), (\ref{astetra1}), (\ref{asPR}) and on the $N_1^*$  
phase factors and weights $(2J+1)$ $\sim(2R)$ in (\ref{standstate}) 
and (\ref{volst}). 
Then the functional which turns out to be associated with the 
resulting configuration is

\begin{eqnarray}
\lefteqn{Z[(T^3,\partial T^3)_{st}
;R \gg 1;L]\,
=}\nonumber \\
& & \Lambda(L)^{-N_0}\,\:\prod_{1}^{N_1-N_1^*} (-1)^{2J} (2J+1)\,
\prod_{1}^{N_3-N_3^E-N_3^V} 
(-1)^{\sum J}\:\ \left\{6j\right\}(J)\nonumber \\
& & \prod_{1}^{N_1^*} (-1)^{2R} (2R)\:\:\Lambda(L)^{-n_0}\:\:
\prod_{1}^{n_1} (-1)^{2j}\:(2j+1)\nonumber \\
& & \prod_{1}^{n_2} \;(2R)^{-1/2}\:
(-1)^{\sum (J+R)+ \sum j+ \Phi}(-1)^{(\sum m)/2}
\left[3j\mu\right]\left[3jm\right]\nonumber \\
& & \prod_{1}^{N_3^E} (2R)^{-1}(-1)^{\sum (J+R)+ j+ \Psi}\:\:
d^j_{\nu_2\,\nu_3}(\theta)\nonumber \\
& & \prod_{1}^{N_3^V} \;(2R)^{-3/2}
(-1)^{\sum (J+R)}\:\exp\left\{i(\sum_{\alpha=1}^{6}l_{\alpha}
\theta_{\alpha} + \pi/4)\right\},
\label{Rstate}
\end{eqnarray}

\noindent where each term containing $(2R)^{-3/2}$ is the approximation for
the corresponding volume factor in (\ref{asPR}).\\  
This expression is an {\em almost factorized} product of two groups of terms\\
{\em i)} A Ponzano--Regge--like
state functional for a $3$-dimensional bulk triangulation to be closed with   
boundary terms depending on some of the $N^*_1$ {\em fixed} edges $J+R$,  
(see below for more details). The sum of this
state functional over all possible triangulations with fixed labels 
on the boundary
(up to regularization) will provide in the bulk a family of $3$-gravity 
partition functions 
depending on the decoupling parameter $R$.\\
{\em ii)} A second functional, represented by the remaining terms in 
(\ref{Rstate}), containing contributions resembling the state 
functional for a $2$-dimensional closed triangulation given in (\ref{bisum}) 
({\em i.e.} a pair of Wigner $3jm$ symbols for each triangular face).\\
Referring to {\em i)}, we recognize that the topological union of the 
coupling tetrahedra, 
$(\{\sigma_F\}\cup\{\sigma_E\}\cup\{\sigma_V\})$, fills in a thick shell 
close to the boundary 
$(\partial{T}^3)_{st}$. The linear extension of such a shell is of the 
order of the decoupling 
parameter $R$, and some of the $N_1^*$ edges considered in (\ref{volst}) 
happen to lie 
on the boundary of the 3-dimensional triangulation

\begin{equation}
{\tilde{T}}^3\:\doteq\:(IntT^3)_{st}\:\setminus\:(\{\sigma_F\}
\cup\{\sigma_E\}\cup\{\sigma_V\}).
\label{newbulk}
\end{equation}

\noindent For each value $R\gg 1$ of the decoupling parameter, 
denote by $\Sigma^{in}(R)$ 
the fixed 2-dimensional triangulation which closes up (\ref{newbulk}) 
giving rise to the pair

\begin{equation}
({\tilde{T}}^3,\, \Sigma^{\,in}(R)).
\label{newpair}
\end{equation}

\noindent Then any such a pair is topologically equivalent to the 
original pair $(T^3,\partial{T^3})_{st}$ as can be easily seen by 
exploiting the invariance under elementary shellings discussed at 
the end of Section 2. Moreover, with respect to $(T^3,\partial{T^3})_{st}$, 
$\Sigma^{\,in}(R)$ represents an {\em inner boundary} with

\begin{equation}
\Sigma^{\,in}(R)\cap(\partial{T}^3)_{st}\:=\:\emptyset .
\label{inters}
\end{equation}

\noindent For what concerns the state functional to be associated 
with (\ref{newpair}), note that only contributions from the edges in 
$\Sigma^{in}(R)$ -- and not from its faces -- have to be taken into 
account. This is due to the fact that
the contribution to the state functional of an edge on a fixed boundary 
amounts simply to $(-1)^{J}
(2J+1)^{1/2}$ \cite{ooguri}, which further reduces to $(-1)^R(2R)^{1/2}$ 
in our case.\\
If we now select 

\begin{equation}
N_1(\Sigma^{in}(R))\:\doteq \: \mbox{number of edges in}\:\Sigma^{in}(R) 
\label{N1in}
\end{equation}

\noindent out of the $N_1^*$ edges in (\ref{volst}), by collecting from 
(\ref{Rstate}) the terms from the bulk, we get

\begin{eqnarray}
\lefteqn{Z[({\tilde T}^3,\Sigma^{in}(R));L]\,
=}\hspace{.5in}\nonumber \\
& & \Lambda(L)^{-(N_0-N_0(\Sigma^{in}))}\:\,\prod_{1}^{N_1 - N^*_1}\; 
(-1)^{2J} (2J+1)\,\prod_{1}^{N_3} 
(-1)^{\sum J}\:\ \left\{6j\right\}(J)  \nonumber \\
& & \prod_{1}^{N_1(\Sigma^{in})} (-1)^{R} (2R)^{1/2}.
\label{bulkst}
\end{eqnarray}

\noindent Notice that here we are forced to include only 
$(N_0 - N_0(\Sigma^{in}))$ vertex weights, $N_0(\Sigma^{in})=$
(number of vertices of the inner boundary), namely those vertices which are
actually in the new $Int {\tilde T}^3$.\\
As remarked before, we have now the possibility of getting at once 
the partition function
of the bulk on applying the standard formal procedure in the case
of a fixed boundary, {\em i.e.}

\begin{equation}
Z[(M^3, \partial M^3\equiv \Sigma^{in}(R)]\,=
\,\lim_{L\rightarrow \infty}\:\:\:
\sum_
{\left \{\begin{array}{c}
Int{\tilde T}^3,J\leq L\\
\Sigma^{in}(R)fixed
\end{array}\right\}}\:
Z[({\tilde T}^3,\Sigma^{in}(R)); L].
\label{bulksum}
\end{equation} 

\noindent This family of Euclidean $3$-gravity partition functions 
encodes information about
the decoupling parameter $R$ in its fixed boundary, while free fluctuations
of all the other spin variables in the bulk are obviously allowed.
In the decoupling limit to be discussed below, the state
sum (\ref{bulksum}) should be sharply peaked on (semi)classical
configurations, namely on states (\ref{bulkst}) in which all spin variables
$J$ in $Int T^3$ have been rescaled by choosing some finite $L\gg1$, 
with $L<R$.
But such conditions do not affect the "topological character"
of (\ref{bulksum}), since Ponzano and Regge showed that also the asymptotic
$6j$ symbols satisfy all the algebraic identities \cite{pr}, and thus the 
bistellar
moves in $Int T^3$ can work as well as before.\\
Coming back now to the factorization of the state functional (\ref{Rstate}),
and taking into account (\ref{bulkst}), we can formally write

\begin{eqnarray}
\lefteqn{Z[(T^3,\partial T^3)_{st};R \gg 1;L]\;=}\hspace {3cm}\nonumber\\
& & Z[({\tilde T}^3,\Sigma^{in}(R)); L]\:\:{\cal P}
(\Sigma^{in},\Sigma^{out}\,;R)\:
Z^{hol}\,[\Sigma^{out};L],
\label{factst}
\end{eqnarray}

\noindent where we have set 
\begin{equation}
(\partial T^3)_{st}\:\equiv\:\Sigma^{out}
\label{outsigma}
\end{equation}

\noindent to denote the (fluctuating) outer boundary, in agreement 
with the notation $\Sigma^{in}$ for the (frozen) inner boundary.\\
The expressions of the last two terms on the right--hand side of 
(\ref{factst}) can be recognized from (\ref{Rstate}), taking into 
account (\ref{N1in}), (\ref{outsigma}) and (\ref{bulkst}). Thus\\

\begin{itemize}
\item In the {\em projection map} 

\begin{eqnarray}
\lefteqn{{\cal P}(\Sigma^{in},\Sigma^{out}\,;R)\:\doteq}\hspace{1cm}\nonumber\\
& & \prod_{1}^{N_1^*-N_1(\Sigma^{in})}(2R)
\prod_{1}^{N_1(\Sigma^{in})}(2R)^{1/2}
\prod_{1}^{n_2}(2R)^{-1/2}\:\prod_{1}^{N_3^E}(2R)^{-1}\:
\prod_{1}^{N_3^V}(2R)^{-3/2}\,=\nonumber\\
& & (2R)^{N_1^*-\frac{1}{2}N_1(\Sigma^{in})}
\left(\frac{1}{2R}\right)^{\frac{1}{2}n_2}\:
\left(\frac{1}{2R}\right)^{N_3^E}\:
\left(\frac{1}{2R}\right)^{\frac{3}{2}N_3^V}
\label{promap1}
\end{eqnarray}

\noindent we collect all the terms of (\ref{Rstate}) which 
depend explicitly from $R$
(the phase factors containing $R$ can be dropped {\em e.g.} 
by choosing an even $R$).

\item The remaining terms of (\ref{Rstate}), not yet considered, are

\begin{eqnarray}
\lefteqn{Z^{hol}\:
[\Sigma^{out}\,;L]\,
=}\hspace{.7in}\nonumber \\
& & \Lambda(L)^{-n_0-N_0(\Sigma^{in})}\:
\prod_{1}^{n_1} (-1)^{2j}\:(2j+1)\nonumber \\
& &   \prod_{1}^{n_2} \:
(-1)^{\sum J+ \sum j+ \Phi}\:\,(-1)^{(\sum m)/2}
\left[3j\mu\right]
\:\,\: \left[3jm\right]\nonumber\\
& & \prod_{1}^{N_3^E}(-1)^{\sum J+ j+ \Psi}\:\:
d^j_{\nu_2\,\nu_3}(\theta)\nonumber \\
& & \prod_{1}^{N_3^V} \;
(-1)^{\sum J}\:\exp\left\{i(\sum_{\alpha=1}^{6}l_{\alpha}
\theta_{\alpha} + \pi/4)\right\},
\label{holostate}
\end{eqnarray}

\noindent where the extra--factor due to the vertices of $\Sigma^{in}$ -- 
which compensates 
the modification of the weights in (\ref{bulkst}) -- has been included. 
This expression
represents the functional to be associated with a holographic state 
generated by disentangling the external boundary from the original 
$3$-dimensional theory.
\end{itemize}

\noindent As a first remark, note that the projection map includes 
combinatorial quantities related 
not only to the topology of the manifold 
$(T^3,\partial{T}^3)_{st}$, but also to the triangulations of $\Sigma^{in}$, 
$\Sigma^{out}$ and of the shell between them. Thus, in particular, 
the numbers $N_1^*$, $N_1(\Sigma^{in})$, 
$n_2\equiv{n}_2(\Sigma^{out})$, $N_3^E$, $N_3^V$ in (\ref{promap1}) 
cannot be related to each 
other in a unique way unless we choose {\em a priori} the topological 
type {\em and} the triangulations. \\
Anyway, we may say that the onset of a {\em decoupling regime} is achieved
when the value of the limit

\begin{equation}
\lim_{R \rightarrow \infty}\:\:{\cal P}(\Sigma^{in},\Sigma^{out};\,R)
\label{limP}
\end{equation}

\noindent is small as compared with the other functionals
appearing in the factorization (\ref{factst}). In other words, we require that
the gravitational contribution
of the shell made up by the coupling tetrahedra is negligible when 
approaching the large 
$R$--limit both with respect to the bulk contribution (\ref{bulkst}) 
and also with respect to the residual functional (\ref{holostate}) 
surviving on the 
external boundary. Such a kind of behavior can be translated into 
suitable sets of
selection rules (or more properly, in 
holographic language, a sort of "screen map" \cite{susskind}) 
on the triangulations involved in (\ref{promap1}).
Indeed, given a topological type $(M^3 , \partial M^3)$, we can in 
principle select -- by making use
of both bistellar moves in the interior of $M^3$ and shellings and/or inverse
shellings in the boundary -- exactly those standard triangulations 
which satisfy the 
constraints, and in the decoupling limit the bulk--boundary transfer 
process is efficiently performed 
only by such classes of triangulations. We shall return to the physical 
interpretation of
this point in the next section.\\
An inspection of the holographic state functional (\ref{holostate}) 
shows that it is not
independent of the triangulation $\Sigma^{out}$.   
This feature is an obvious consequence of the factorizazion 
prescription (\ref{factst}),
 which breaks
the topological invariance of the original 
$Z[(T^3, \partial T^3)_{st}$ $\rightarrow
(M^3,\partial M^3);$ $\,L]$. Even if such an
invariance can be restored in the bulk according to the procedure 
given in (\ref{bulksum}),
by comparing (\ref{holostate}) with the bidimensional 
functional (\ref{bistate}) we
see that in the present case a number of new significant 
quantities appear (the phase factors are 
unimportant since we could redefine in a suitable way the Wigner symbols). 
To address this issue in more details, recall from Section 2 and 
from \cite{carbone3} that the 
$2$-dimensional bistellar moves are

\begin{itemize}

\item The {\em flip move}, which tranforms a pair of contiguous triangles
$\tau_1,\tau_2$  into another pair of triangles $\tau'_1,\tau'_2$ by 
keeping their
common boundary quadrilateral fixed. Denote this move by      

\begin{equation}
[\tau_1\,,\tau_2\,] \longrightarrow [\tau'_1\,,\tau'_2\,].
\label{topflip}
\end{equation}

\noindent The algebraic identity representing this move is

\begin{eqnarray}
\lefteqn{\sum_q \sum_{\kappa,\kappa'}
(2q+1)(-1)^{2q}\,(-1)^{-\kappa -\kappa'}
\left[ \begin{array}{ccc}
p & a & q \\
\psi & \alpha & -\kappa
\end{array}\right]_{\tau_1}
\left[ \begin{array}{ccc}
q & b & r \\
\kappa & \beta & \rho
\end{array}\right]_{\tau_2}}\nonumber\\
& & \left[\begin{array}{ccc}
p & a & q \\
\psi' & \alpha' & -\kappa'
\end{array}\right]_{\tau_1}
\left[ \begin{array}{ccc}
q & b & r \\
\kappa' & \beta' & \rho'
\end{array}\right]_{\tau_2}
\,=\,\sum_c 
\sum_{\gamma, \gamma'} (2c+1)\,(-1)^{2c}\,(-1)^{-\gamma -\gamma'}\nonumber\\
& &  \left[ \begin{array}{ccc}
a & b & c \\
\alpha & \beta & \gamma
\end{array}\right]_{\tau'_1}
\left[ \begin{array}{ccc}
r & p & c \\
\rho & \psi & -\gamma
\end{array}\right]_{\tau'_2}
\left[ \begin{array}{ccc}
a & b & c \\
\alpha' & \beta' & \gamma'
\end{array}\right]_{\tau'_1}
\left[ \begin{array}{ccc}
r & p & c \\
\rho' & \psi' & -\gamma'
\end{array}\right]_{\tau'_2}
\label{flipid}
\end{eqnarray}

\noindent where Latin letters denote angular momentum variables, 
Greek letters the first set
of momentum projections and primed Greek letters the second one. 
A label associated with
the triangles involved in the move has been added.

\item The {\em Alexander} (or barycentric) {\em move} which amounts to adding 
one vertex in the interior 
of a triangle $\tau$ giving rise to three triangles 
$\tau'_1$, $\tau'_2$, $\tau'_3$
bounded by the original one. We denote this operation and its inverse move by

\begin{equation}
[\tau\,] \longleftrightarrow [\tau'_1\,,\tau'_2\,,\tau'_3]
\label{alex}
\end{equation}

\noindent and the corresponding identity reads

\begin{eqnarray}
\lefteqn{\sum_{q,r,p} 
(2q+1)(2r+1)(2p+1)\,(-1)^{2q+2r+2p}
\sum_{\kappa,\kappa'}\sum_{\rho,\rho'}\sum_{\psi,\psi'}
(-1)^{-\kappa -\kappa'}(-1)^{-\rho -\rho'} }\hspace{.5in}\nonumber\\
& &  (-1)^{-\psi -\psi'} \left[ \begin{array}{ccc}
p & a & q \\
\psi & \alpha & -\kappa
\end{array}\right]_{\tau'_1}
\left[ \begin{array}{ccc}
q & b & r \\
\kappa & \beta & -\rho
\end{array}\right]_{\tau'_2}
\left[ \begin{array}{ccc}
r & c & p \\
\rho & \gamma & -\psi
\end{array}\right]_{\tau'_3} \nonumber\\
& & \left[ \begin{array}{ccc}
p & a & q \\
\psi' & \alpha' & -\kappa'
\end{array}\right]_{\tau'_1}
\left[ \begin{array}{ccc}
q & b & r \\
\kappa' & \beta' & -\rho'
\end{array}\right]_{\tau'_2}
\left[ \begin{array}{ccc}
r & c & p \\
\rho' & \gamma' & -\psi'
\end{array}\right]_{\tau'_3}\,=\nonumber\\
& & \Lambda(L)^{-1}\,
\left[ \begin{array}{ccc}
a & b & c \\
\alpha & \beta & \gamma
\end{array}\right]_{\tau}
\left[ \begin{array}{ccc}
a & b & c \\
\alpha' & \beta' & \gamma'
\end{array}\right]_{\tau}
\label{baryid}
\end{eqnarray}

\end{itemize}

\noindent By comparing the content of the above identities with the 
expression of the
holographic state functional (\ref{holostate}) we observe that

\begin{itemize}

\item The asymptotic value of the $6j$ symbol corresponding to a 
tetrahedron of type $\sigma_E$
(see (\ref{astetra1})) provides a $d$-function which decorates the 
edge $j$ in $\Sigma^{out}$.
Thus, if we consider a pair of triangular faces sharing this edge, 
we cannot perform on them the flip move
(\ref{topflip}) anymore.

\item From the asymptotic value (\ref{astetra}) for the $6j$ symbol 
of type $\sigma_F$
we get a decoration of a face in $\Sigma^{out}$. Then the barycentic 
move $[\tau\,] \rightarrow [\tau'_1\,,\tau'_2\,,\tau'_3]$ is forbidden 
in such a circumstance.

\item Finally, the asymptotics of the $6j$ symbol of type $\sigma_V$ 
generates a decoration
of a vertex in $\Sigma^{out}$. As a consequence of that, the inverse 
barycentric move
$[\tau'_1\,,\tau'_2\,,\tau'_3]$ $\rightarrow$ $[\tau\,]$ turns out to 
be forbidden too.

\end{itemize}

\noindent Summing up, none of the moves represented by the identities 
(\ref{flipid}) and (\ref{baryid}) can be freely performed
on the functional (\ref{holostate}) and thus, owing to \cite{pachner1}, 
the topological invariance
cannot be restored in the holographic theory.

\section{Comments and remarks on the holographic theory}
We now examine in more detail some aspects of the 
boundary theory obtained and suggest possible lines 
of further  investigations.\\
First of all we find agreement with the holographic principle in general.
Indeed the boundary answer is expected to depend on the asymptotic
properties of the manifold. In other words, one is  
examining the asymptotic behaviour of the bulk fluctuations
in a region close to the boundary and according to their behaviour one may 
have different projections. It was guaranteed that the topological 
invariance of the theory would be broken by our limit as the limit
basically freezes the standard triangulation as special thus breaking the
symmetry between this configuration and all the other configurations 
that were obtainable by shelling moves. \\
Notice however that the theory is not as far from a topological
theory as it may seem. In fact if we had included non-trivial 
fixed connection in the boundary theory (which indeed is necessary in 
the end for our path sum to have correct gluing properties) then we would
have found that every edge of the dual graph on the boundary would carry 
a Wigner function corresponding to the connection on that edge 
\cite{ooguri,martin} and thus of the same form as the Wigner $d$ function 
that we find from the tetrahedra $\sigma_E$ in (\ref{holostate}). 
In the discussion of 
\cite{ooguri} this is very similar to the squashed partition functions that
are shown to be topological invariant two-dimensional path-sums. 
In our case the various additional factors that accompany this sum 
break this topological invariance, and in addition the Wigner $d(\theta)$ 
depend on the asymptotics of the manifold, whereas the Wigner functions
that come from the boundary conditions depend 
upon the boundary geometry only. The dependence of $d(\theta)$ on asymptotics 
is very suggestive of the derivation used in \cite{BHliou} where 
the asymptotics on connections is the key to reducing a topological
boundary theory to Liouville theory. 
\\
In the case of AdS/CFT this dependence on asymptotic geometry turns 
out to be crucial when evaluating the
boundary correlators. But also in the original 't Hooft
formulation this ``near horizon'' region is precisely the one in which
one has interactions and ultimately correlations between ingoing and
outgoing observables describing the black hole. Again the geometry of the
black hole which  induces 
violent boosts when a particle approaches the horizon, is 
responsible for these high energy scatterings whose phase shifts build up
the hologram on the horizon.\\
As pointed out in \cite{martin} this
asymptotic dependence turns out to be a general feature when
implementing decoupling limits 
in the Ponzano-Regge set up. In particular the configuration considered
in the previous section can be interpreted as follows: once the scale R
is introduced one 
has a family of regions labelled by R inside the bulk which can be thought
of as a sort  of internal 
fluctuating nuclei  and  at the same time  one is isolating
the asymptotic dependence of the bulk
fluctuations by progressively integrating out a region of size R
containing tetrahedra which act as messengers of bulk
information.\\
The boundary will behave as a screen
storing bulk information depending on the value of
what has been called the projection map. Note however that one can always
say (independently of the value of the screen map) that the
dominant contribution
will be given by those tetraedra having a face on the boundary, followed
by the ones which have an edge on the boundary and finally by those with
only one vertex. This follows from the power law $R$ decay previously
considered in the asymptotic formulae for the coupling tetraedra and gives
a simple criterium to examine the spectrum of fluctuations of the
boundary theory.\\
The bulk dependence of the boundary partition function is intimately
related to the asymptotic behaviour of bulk fluctuations. One has
in other words a sort of dressing of the boundary theory induced by the
gravitational bulk fluctuations. This is in some sense similar to the 
AdS/CFT case  where the source terms for the boundary theory couple
a boundary operator to the asymptotic behaviour of a bulk field. In other 
words  one might say that the bulk theory only sees, through the
asymptotic behaviour of its 
fields, the ``abstract'' boundary theory and not its elementary fields.\\
In particular the boundary functional (\ref{holostate}) can be associated 
with a fat trivalent  graph (dual to the $2$-dimensional triangulation) 
containing
a pair of $3jm$ symbols for each vertex and Wigner functions connecting
them.  Now it is a classical result that the Teichm\"{u}ller
space of a Riemann surface can be combinatorially described in terms of 
trivalent ribbon
graphs. So our boundary theory seems to have the right structure in order 
to construct a Teichm\"{u}ller space though a precise link
between our colourings and the moduli of the surface still needs to
be found. Let us also point out that the Teichm\"{u}ller space
of the 2d boundary theory should be indeed related to Einstein gravity
in the three dimensional spacetime \cite{herman1}.\\
The boundary theory is then constructed by multiplying together  
$3jm$ symbols associated with $SU(2)$ representations. On the other hand
according to standard arguments we expect to have on the boundary 
some kind of WZW theory and eventually Liouville theory. 
As a preliminary step it would  then be
interesting to have a dictionary for our boundary theory formulated
in terms of  group theory language, from a CFT perspective in a similar 
spirit as the dictionary
proposed by Moore and Seiberg \cite{sm} between group theory and
classical CFT. They show that to a compact group G there 
corresponds a chiral algebra and to 
Clebsch--Gordan coefficients correspond 
chiral vertex operators. Invariant tensors built
up in group theory correspond to the conformal blocks of the CFT. In addition
Racah coefficients are associated with the fusion matrix and in particular
for the case of $SU(2)$ its matrix elements are nothing 
else than $6j$ symbols. 
Functions on the group are then the physical fields of the CFT and the product 
of functions on the group corresponds therefore to the OPE. The average of
this product represents the physical correlation functions. 
Therefore the product of $3jm$ symbols is related to the product of matrix
elements of chiral vertex operators and according to the BPZ program it would
then be possible to determine all physical quantities of the CFT. 
We find it intriguing that the 
boundary theory that is represented by our rescaled path-sum may be related 
by a suitable dictionary to an (as yet unspecified!) CFT.\\
Of course to get a precise definition of the boundary theory one should 
consider the dual  model living in the bulk Anti-de Sitter three
dimensional spacetime. The natural candidate  would be the 
Turaev-Viro partition sum being a discretized gravity model which 
accomodates for the presence of a non vanishing cosmological constant. 
This model would also provide a well defined partition sum function 
being a regularized version of the Ponzano-Regge model\cite{turaev,mizo}.
However it is not clear how a decoupling procedure of the kind 
proposed in Section 3  could be implemented in such a context as the 
quantum group spin labels are strictly bounded above and maybe another 
approach is necessary.\\ 
A preliminary question is indeed to select the relevant representations
to be summed over in the partition function of the bulk model. Now it 
appears that 
in order to implement conformal bootstrap program in LFT the algebra of 
conformal blocks in LFT has to be related to the algebra
of irreps of a certain quantum group \cite{teschner}. A natural guess is then
to construct a partition function in which one sums over these classes of 
representations \cite{martin,krasnov}.\\
 According to some recent analysis  
\cite{teschner} these representations form a series of continuous irreps of 
$SL_q(2,R)$ with the deformation parameter 
\be
q=\exp (\pi i b^2),
\ee
$b \in (0,1)$ and irrational.
 This series has a very important property: it closes under taking tensor 
products of reps and this would assure the triangulation
independence of the bulk model.\\
 However the bulk model would not be automatically guaranteed to be finite 
since one has first to give sense to the integrals
over these continuous series of representations. In other words one is faced 
up with the well known problem of constructing simplicial partition 
functions in
the case in which the group is non compact. In addition, this series of 
representations turns out not to have a classical limit, corresponding to the 
Ponzano-Regge model.\\
As an aside remark one could consider
the case of
hyperbolic 2+1 dimensional manifolds, having in mind the AdS case to get a 
concrete correspondence with LFT. One is then led to the notion of the
so called hyperbolic knot, a knot that has a component that can be given a 
metric of costant negative curvature. Now the volume of the complement is a 
topological invariant, called the hyperbolic
volume of a knot. In particular one can show that the quantum dilogarithm can 
lead to a generalized deformed notion of the hyperbolic volume. This seems 
to imply that a simplicial TQFT defined by the quantum dilogarithm can in 
principle
be associated to 2+1 dimensional quantum gravity. We refer to 
\cite{kashaev1} and corresponding references for a discussion of these 
points.\footnote{We also point
out that the same author has been relating the quantum hyperbolic invariant of
knots to the quantum theory of Teichm\"{u}ller spaces of punctured surfaces 
\cite{kashaev2}.}\\
Finally we would like to note that our example here is only three-dimensional
and due to the topological nature of gravity in three dimensions is 
somewhat special as all degrees of freedom necessarily reside on the boundary. 
This model is useful however in the context of studying holography 
in discretized gravity as it provides an indication of 
how to proceed to study holography in discretized gravity in higher 
dimensions (for a review of higher dimensional discretized 
gravity see \cite{rwilliams}). Furthermore, the eventual
derivation of discretized Liouville theory using our methods would be 
a very interesting and non-trivial result in its own right. 


\section{Acknowledgements}
G.A. would like to thank E.Martinec for a pleasant discussion during 
the Amsterdam Summer Workshop on String Theory. M.O'L would like
to thank M. Blau and G. Thompson for various topological conversations. 
G.A. and M.O'L. are supported by the EEC under RTN
program HPRN-CT-2000-00131. G.A. is also partially supported by MURST.
M.C. and A.M. are partially supported by MURST, under PRIN grant 9901457493
({\em Geometry of Integrable Systems}). \\

\end{document}